\documentclass[aps,prl,reprint,groupedaddress]{revtex4-1}
\usepackage{graphicx,color}
\usepackage{amssymb}   
\usepackage{amsmath,ulem}
\usepackage{epstopdf}
\usepackage[hidelinks]{hyperref}
\usepackage{bm}
\usepackage{cancel}

\begin{document}

\title{Optical Responses of Chiral Majorana Edge States in Two-Dimensional Topological Superconductors}
	
\author{James Jun He$ ^1 $, Yukio Tanaka$ ^2 $, Naoto Nagaosa$ ^{1,3} $}
	
\affiliation{
		$ ^1 $ RIKEN Center for Emergent Matter Science (CEMS), Wako, Saitama 351-0198, Japan\\	
		$ ^2 $ Department of Applied Physics, Nagoya University, Nagoya 464-8603, Japan\\
		$ ^3 $ Department of Applied Physics, The University of Tokyo, Tokyo 113-8656, Japan
	}
	
\date{\today}

\begin{abstract}
Majorana fermions exist on the boundaries of two-dimensional topological superconductors (TSCs) as charge-neutral quasi-particles. The neutrality makes the detection of such states challenging from both experimental and theoretical points of view. Current methods largely rely on transport measurements in which Majorana fermions manifest themselves by inducing electron-pair  tunneling at the lead-contacting point. Here we show that chiral Majorana fermions in TSCs generate {enhanced} local optical response. The features of local optical conductivity distinguish them  not only from trivial superconductors or insulators but also from normal fermion edge states such as those in quantum Hall systems. Our results provide a new applicable method  to detect dispersive Majorana fermions and may lead to a novel direction of this research field. 
\end{abstract}

\maketitle

The detection and manipulation of Majorana fermion in solids is a keen issue from the viewpoints of both fundamental physics and applications \cite{Kitaev2001,Wilczek}. 
Majorana fermions are neutral and almost free from interactions.
Therefore, it is a challenge to observe and explore experimentally the physical consequences of the Majorana fermions in solids. Topological superconductors (TSCs) \cite{Sato,Tanaka_JSPS} are regarded as promising candidates to realize Majorana fermions, where the Majorana bound state appear at the core of the vortex under magnetic field or the propagating Majorana edge channel exists at the boundary of the sample \cite{Read,Fu2008,Qi2010,Chung2011,JJH_CP2019}. 

Scanning tunneling spectroscopy (STM) is a powerful tool to detect the Majorana bound state at the zero energy \cite{Vic2009,Lunhui,Chiu,JiaPRL,Yazdani2014,Machida}. However, there are other possible reasons for the bound states near zero at the core of the vortex, and it is difficult to exclude these other possibilities. Recent advances
are the detection of the quantized conductance $ G=(2e^2)/h $ \cite{Kouwenhoven2018}, and the high resolution STM at low temperatures \cite{Machida}. On the other hand, the propagating Majorana edge channel is less investigated. The half-quantization of the conductance, $ G=e^2/2h $ , in the structure made of quantized anomalous Hall system and superconductor on the surface of topological insulator was proposed to be an 
	evidence for this Majorana edge channel \cite{Chung2011,JWang2015,Qinglin2017}.	
However, other possible reasons to explain the half-quantization of the conductance were proposed \cite{Wen2018,Sau2018}, and hence the situation is not convincing yet. STM could be useful also for the Majorana edge channel \cite{ZWang}, but the energy dispersion gives the finite density of states for the local probe, not the sharp peak. Therefore, it is desired to explore the spectroscopy of the Majorana edge channel more in depth. 

Microwave spectroscopy with spatial resolution has been applied to the quantum Hall systems \cite{Lai} and recently also to the quantized anomalous Hall systems \cite{Shen2}. The low frequency optical conductivity and dielectric function can be detected as functions of spatial position, and the response of the chiral edge channel has been successfully observed. This is reasonable because the chiral edge channel is gapless and metallic. Majorana edge channel is, on the other hand, neutral and hence naively does not respond to the electromagnetic field. This can be understood from the identity $ \gamma=\gamma^\dagger $and $\gamma^\dagger \gamma=\gamma^2 = constant$ for the creation and annihilation operators of Majorana fermion. Therefore, the continuity equation of charge appears to require $ \nabla \cdot \bm J =0 $ ($ \bm J $: current density) which prohibits current response in one dimension. 
	However, the exchange of charge between the edge and bulk occurs
as has been discussed in quantum Hall system \cite{NN1995}, topological insulator \cite{Fukui}, and topological superconductor \cite{Fukui}.

In this paper, we investigate the local optical conductivities, $ \sigma_{xx}(\omega) $, of two-dimensional (2D) TSCs, especially of their Majorana edge channels. We start with a model of spinless $ p+ip $ SC where the Majorana edge modes generate a frequency-dependent signal. This is followed by a 1D effective model analysis, which provides a clear physical picture about the origin of the Majorana signals and predicts the frequency dependence $\sigma_{xx}(\omega) \sim \omega^2 $. Such a result is in sharp contrast to the $ \omega $-independent one of normal edge modes. The optical signals across the topological phase transitions are studied with a model of quantum anomalous Hall (QAH) insulators in proximity to a SC.

\paragraph{ 2D p+ip-superconductors ---}

\begin{figure}
	\includegraphics[width=0.9\linewidth]{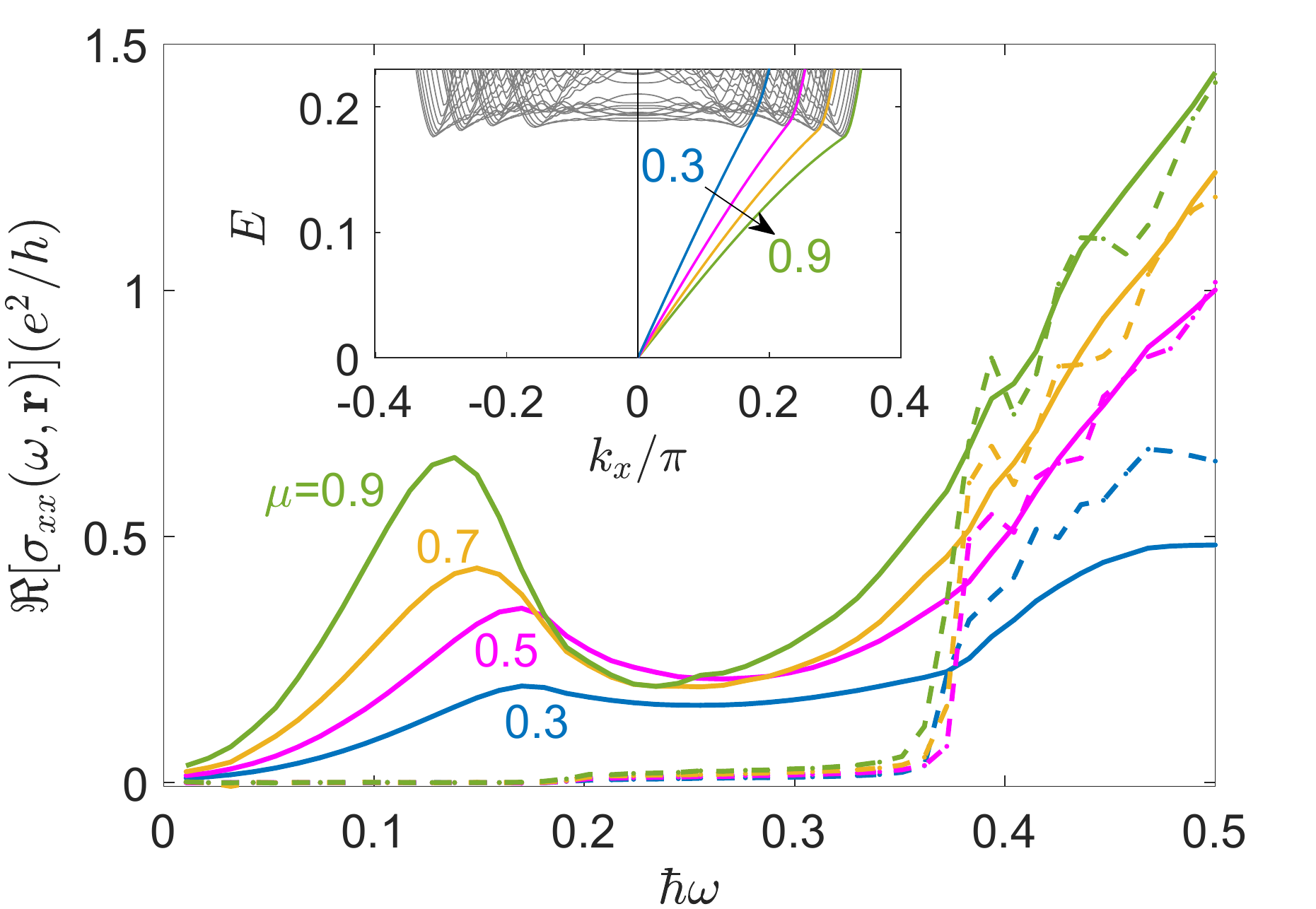}
	\caption{The frequency dependence of local longitudinal optical conductivity on the edge of a spinless p+ip superconductor for various values of chemical potential $ \mu $. The dashed  curves are calculated at a region in the bulk. 
		The inset shows the evolution of the edge states as $ \mu $ varies. 
		Other parameters: lattice size along y-direction $ L_y=60 $, hopping $ t=1 $ is used as the energy unit, pairing amplitude $ \Delta=0.2 \sqrt{t/\mu} $, temperature $ k_B T=0.001 $. The spot size is $ X=1 $ and $ Y=4 $. }
	\label{fig:pwavemus}
\end{figure}

Consider a 2D spinless $ p+ip $ superconductor described by the following tight-binding Hamiltonian,
\begin{eqnarray}
H_\text{p+ip} = &&  \sum_{\bm r}   \sum_{{\bm d}=\pm\hat{\bm x},\pm\hat{\bm y}} [-t \psi_{\bm r}^\dagger \psi_{\bm r+{\bm d}}  + (\Delta_{\bm d} \psi_{\bm r}^\dagger \psi_{\bm r+\bm {d}}^\dagger +h.c.)] \notag 
\\
&& - \sum_{\bm r} (\mu-4t)  \psi_{\bm r}^\dagger \psi_{\bm r} ,
\label{eq:tbp}
\end{eqnarray}
where the summation $ \sum_{\bm r} $ is over the sites on a two-dimensional square lattice which is infinite in the $ x $-direction but finite along the $ y $-direction. $ t $ is the hopping, $ \mu $ is the chemical potential and  $ \Delta_{\bm d} $ is the pairing between neighbouring sites, given by $ \Delta_{\pm \hat{\bm x}}=i \Delta_{\pm \hat{\bm y}}=\pm \Delta $ where $ \hat{\bm x} $ and $ \hat{\bm y} $ are unit vectors along $ x $- and $ y $-directions respectively.  Quasi-1D chiral Majorana states appear on the edges when $ 0<\mu<8t $. The $ x $-direction current density operator is $ j_x(\bm r) = i (e t/\hbar) (\psi_{\bm r}^\dagger \psi_{\bm r+\hat{\bm x}}  - h.c. )$ and the current operator in a finite region $ 0\leq x\leq  X  $ and $ 0 \leq y \leq Y $ is
\begin{eqnarray}
J_x(\bm r) = \frac{1}{X}\sum_{m=0}^{X}\sum_{n=0}^{Y} j_x(\bm r+m \hat{\bm x}+n \hat{\bm y}).
\end{eqnarray}
Assuming that the light only shines on this region, one obtains the optical conductivity
$
\sigma(\omega,\bm r) =\omega ^{-1}\int_0^\infty dt  e^{i\omega t} \langle [{J}_x(\bm r,t),{J}_x(\bm r,0)] \rangle , 
$
where $ \omega $ is the photon frequency 
 and this formula is calculated using the Green's functions.\cite{supp}

The real-part conductivities for various values of chemical potential $ \mu $ are shown in FIG. \ref{fig:pwavemus}, with the temperature $ T=0.001/k_B $ (the $  T$-dependence is shown in \cite{supp}) and the spot size $ X=1,Y=4 $.
From the bulk values of $ \Re[\sigma_{xx}(\omega,\bf r)] $, i.e. the dashed curves in FIG. \ref{fig:pwavemus}, we can tell the optical gap is $ 2\Delta_g \approx 0.37$ and thus the bulk superconductivity gap $ \Delta_g \approx 0.18 $. 
The pairing amplitude varies with $ \mu $ so that the bulk gap keeps approximately unchanged. The $ \omega $-dependence of $  \Re[\sigma_{xx}(\omega,\bf r)] $ shows a peak near $ \hbar\omega \approx \Delta_g $. 
At energies higher than the peak position, the curves start to increase again due to the joining of the bulk states when $ \hbar\omega >\Delta_g $. 
That is, a photon may create an in-gap Majorana state along with a bulk state whose energy is larger than $ \Delta_g $, giving a contribution to the optical conductivity.
For $ \hbar\omega>2\Delta_g $, the main contribution comes from the bulk and thus the positions in the bulk and on the edge give similar results. The results with different values of the chemical potential $ \mu $ are similar. 
But the magnitude in the small frequency regime increases as we increase $ \mu $ since the dispersion becomes flatter, giving a larger density of states, as shown in the inset of FIG.  \ref{fig:pwavemus}.

\paragraph{ 1D analysis --- }

To understand the origin and the features of the optical conductivity contributed by the Majorana edge modes, it is helpful to do a 1D analysis with the effective edge Hamiltonian

\begin{align}
H_{\text{eff}} =-iv \int_0^L dx \gamma(x) \partial_x \gamma(x).
\label{eq:h0}
\end{align} 
where $ \gamma^\dagger(x)=\gamma(x) $ is the edge Majorana field operator. 
For convenience, we shall rewrite it in reciprocal space using the transformation
$
\gamma(x) = \frac{1}{\sqrt{L}}\sum_{k>0} [\gamma_k e^{ikx} + \gamma_k^\dagger e^{-ikx}], 
$
where 
 $ L $ is the length of the hypothetical 1D system. 
 The Hamiltonian becomes
\begin{eqnarray}
H_{\text{eff}} =\sum_{0<k<\Delta/v} v k \gamma_k^\dagger \gamma_k, 
\end{eqnarray}
where we apply the energy cut-off $ \Delta $ which can be regarded as the bulk energy gap of a TSC. 
 
\begin{figure}
	\centering
	\includegraphics[width=0.8\linewidth]{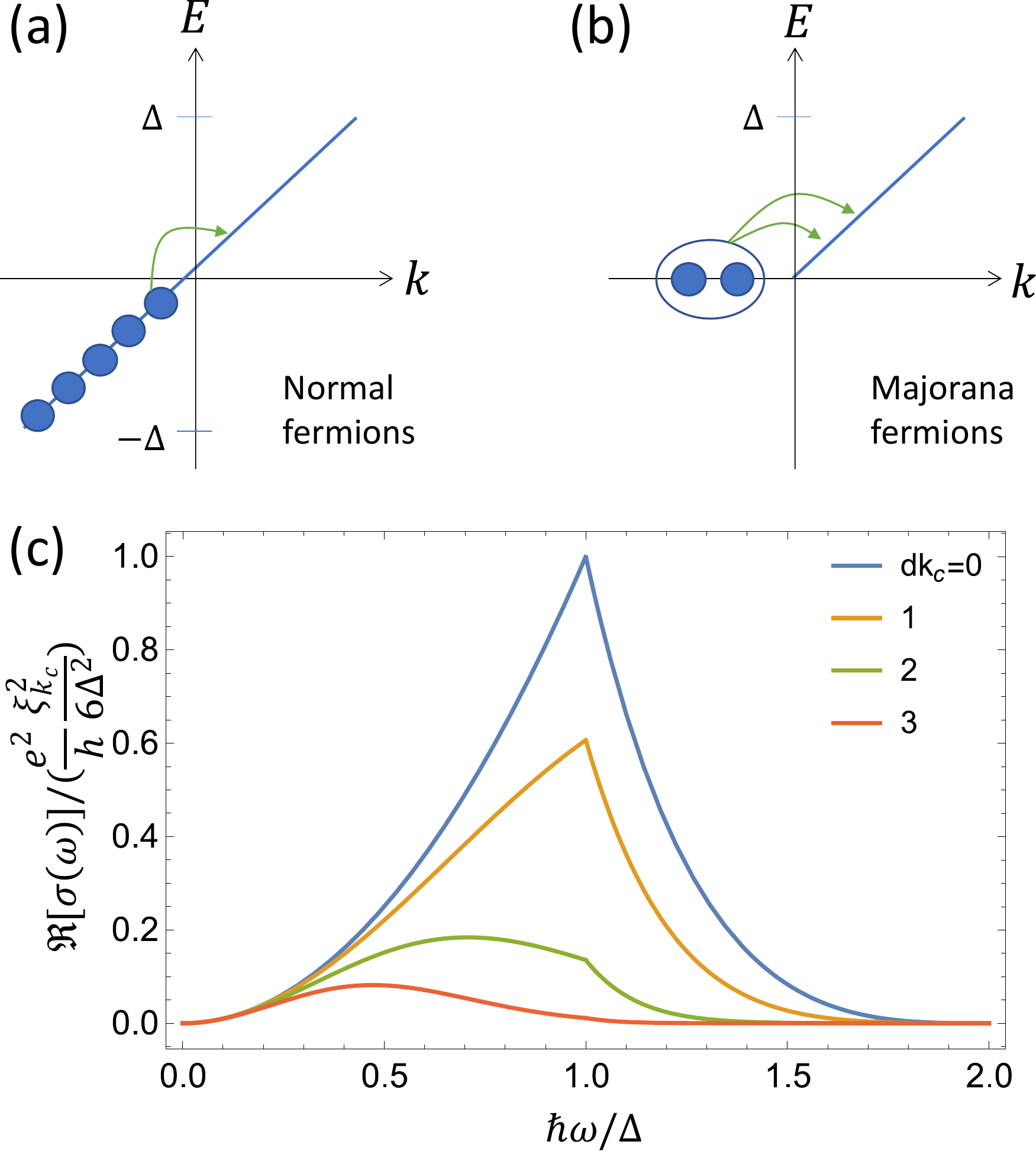}
	\caption{ (a) Schematic energy spectrum of chiral normal fermions. A photon can excite an electron in the Fermi sea to a state above the Fermi energy when the excitation is local, i.e., the momentum conservation is broken. (b) For Majorana fermions, the wave vector is limited to $ k>0 $.
		A photon can create a pair of Majorana fermions from a Cooper pair. (c) The real part of optical conductivity $ \Re[\sigma(\omega)] $ due Majorana states shown in (b). 
	}
	\label{fig:fig1}
\end{figure}
 
For normal chiral fermions, the ground state is obtained by occupying the states below the Fermi energy. Photon absorption happens by exciting electrons to higher-energy empty states, as shown in FIG. \ref{fig:fig1}(a). In contrast, the ground state of a TSC with chiral Majorana modes consists of a Cooper pair condensate and the absorption of photons breaks Cooper pairs into Majorana modes, as shown in FIG. \ref{fig:fig1}(b). The resulting optical conductivity is
$
\sigma(\omega,q) =(\omega L)^{-1}\int_0^\infty e^{i\omega t} dt \langle [J^\dagger(q,t),J(q,0)] \rangle , 
$
where
\begin{align}
J(q)  =  \frac{e \hbar }{ m^* }   \sum_{k>0} [& (q+2k) \theta_{q+k} \gamma_{k+q}^\dagger \gamma_{k}  + (\frac{q}{2}-k) \theta_{q-k} \gamma_{q-k}^\dagger  \gamma_{k}^\dagger
\notag \\
&+(q/2+k) \theta_{-(q+k)} \gamma_{-(q+k)} \gamma_{k} ],
\label{eq:current}
\end{align}
is the edge current operator obtained by projecting the bulk one onto the Majorana edge states.\cite{supp} The parameter $ m^* $ is the effective mass which depends on the specific system. In chiral $ p $-wave SCs, $ m^* $ is just equal to the bulk electron mass \cite{Furusaki}.

At $ T=0 $, the real part of the optical conductivity is
\begin{eqnarray}
\Re[\sigma(\omega,q)] =&& \frac{e^2}{6 \hbar} \frac{(\hbar\omega \xi_{k_c})^2}{\Delta^4} \delta(q-\hbar\omega/v) 
\notag
\\
&&\times \left\{ 
\begin{array}{ccc}
1,     &&\hbar\omega/\Delta \in [0,1], \\
\left(\frac{2\Delta}{\hbar\omega}-1\right)^3 ,  &&\hbar\omega/\Delta \in [1,2], \\
0 ,  &&\hbar\omega/\Delta > 2.
\end{array} 
\right. 
\label{eq:sigma0}
\end{eqnarray} 
where $ k_c =  \Delta/v $ and $ \xi_k =  \frac{\hbar^2 k^2}{2m}  $.
To relate to optical microscopy measurements, let us assume the detecting light to have a Gaussian-distributed  intensity  $ g(x)= \exp(-x^2/d^2)$. Then, the quantity detected with optical microscopy methods is
\begin{eqnarray}
\sigma(\omega) =&& \frac{1}{\pi d^2} \int dx \int dx' g(x) g(x') \sigma(\omega;x-x') ,
\\
= && [\sigma(\omega)]_{d=0} \exp\{-\frac{(dk_c)^2}{2}  \frac{(\hbar\omega)^2}{\Delta^2}\} .
\label{eq:sigma1}
\end{eqnarray}
The real part of $ [\sigma(\omega)]_{d=0} $ is simply given by the right-hand-side of Eq.(\ref{eq:sigma0}) with the delta function omitted. 

In FIG. \ref{fig:fig1}(c), $\Re[ \sigma(\omega)] $  at $ T=0  $ is shown for various values of light distribution width $ d $. 
The real part of zero-temperature optical conductivity vanishes at $ \omega=0 $ and increases quadratically for small $ \omega$. As the frequency becomes large, the number of processes to absorb the photon with frequency $ \omega $ starts to decrease and thus $\Re[ \sigma(\omega)] $ shows a peak.  When $ \hbar\omega>2\Delta $, the photon energy exceeds the sum of any two Majorana fermions' and thus absorption cannot happen, yielding zero $ \Re[\sigma(\omega)] $. 
Although the static ($ \omega=0 $) local conductivity vanishes at $ T=0 $, it is non-zero when $ T>0 $. 
In fact, for $ k_B T \ll \Delta $, 
considering the Fermi distribution of the quasi-particles, we obtain the on-site ($ d = 0 $) response  
$
\Re[\sigma(\omega=0;T)_{d=0}] =   \frac{e^2}{3h} \frac{ \xi^2_{k_c}}{\Delta^4} (\pi k_B T)^2.
$

When $ d=0 $, the peak is at $ \hbar\omega=\Delta $. The sharpness of the peak comes from the abrupt energy cut-off  assumed.
If the light shines on a finite region and thus $ d  $ is increased, the signal is reduced because the translational symmetry is gradually recovered and the transition between states with different momenta is suppressed. 
Also, the peak position of $\Re[ \sigma(\omega)] $ is shifted towards lower frequencies since larger $ d $ means that low-wave-vector (low-energy) states contribute more than high-wave-vector ones
and thus the low-frequency response is enhanced relatively.  
For large $ d k_c $, the peak position according to Eq. (\ref{eq:sigma1}) is at $ \hbar \omega_0/\Delta = \sqrt{2}/(d k_c) $ where the peak height is 
$
\Re[\sigma(\omega_0)_{dk_c\gg 1}] = \frac{e^2 }{h} \frac{\xi_{k_c}^2}{3\Delta^2} \frac{\exp\{-1\}}{(dk_c)^2}. 
$
Thus the signal decreases as $ \sim d^{-2} $ when $ d $ increases. 

As we have seen, the 1D analysis is entirely consistent with the previous numerical results of the 2D $ p+ip $ TSC. It clarifies the origin of the optical conductivity and the reasons behind the frequency dependences throughout the sub-gap regime.  Furthermore, the 1D results are helpful to a realistic estimation of Majorana-mode-induced $ \Re[\sigma_{xx}(\omega)] $, since they apply to various systems.

For example, assume the Majorana fermions to be the edge states of a 2D TSC with an energy gap $ \Delta \approx 10^{-4} \mu $, $ \mu $ being the chemical potential. Then $ k_c $ is basically  the Fermi wave vector $ k_F $ and thus $ \xi_{k_c} \approx  \mu$. If  $ k_c \approx k_F \approx 1 \AA^{-1}  $ and $ d \approx 1\mu m $  so that $ d k_c \approx 10^4 $. 
The peak position in the case of chiral Majorana modes is $ \hbar \omega_0/\Delta  \approx  10^{-4} $ and the peak height is about $ 0.1e^2/h $, comparable to the $ 0.5e^2/h $ of the chiral normal fermions. 

If $ d $ is reduced by one order of magnitude ($d \sim 0.1\mu m $), the Majorana signal is enhanced by 100 times, becoming much larger than in the normal case. Such strong optical response of Majorana fermions is due to the large density of states $ \mathcal{N} $. 
Assuming linear dispersion, we have $ \mathcal{N}= 1/v = k_F/\Delta $. With given $ k_F $, small gap $ \Delta $ indicates a small $ v $,  giving a large $ \mathcal{N} $. For normal chiral fermions, the current operator is proportional to $ v $ and thus the enhanced density of states is compensated by the reduction in the current operator, resulting in the constant conductivity. However, for Majorana fermions, $ J \propto  \frac{\hbar k}{m} $  (Eq.(\ref{eq:current})) does not depend on $ v $. The enhancement in $ \mathcal{N} $ is not compensated and the optical response becomes large. 
Recently, a microwave microscopy experiment with an ultrahigh spatial resolution of $ 5nm $ has been reported. \cite{Lee_SciAdv2020} With this size used for $ d $, $ \Re[\sigma_{xx}] $ achieves $ 4000 e^2/h $. 

For comparison, the optical conductivity 
of normal chiral fermions (as illustrated in FIG.\ref{fig:fig1}(a)) can be calculated in a similar way and it is a constant
 $ \Re[\sigma^N(\omega)]= e^2/2h $ when $ \hbar \omega< \Delta$ and $ k_B T \ll \Delta $.

\paragraph{ Transition between QAH and TSC ---}
Consider the following 2D Hamiltonian \cite{Qi2010},
\begin{eqnarray}
H = &&\sum_{\bm k} \psi_s^\dagger(\bm k) [M(\bm k)\sigma^z_{ss'} + A (\bm k\cdot {\bm \sigma})_{ss'} -\mu \delta_{ss'}]  \psi_{s'}(\bm k) \notag
\\ 
+ &&\sum_{\bm k}   \psi_s^\dagger(\bm k) \Delta_\text{sc} i\sigma^y_{ss'} \psi_{s'}^\dagger(-\bm k),
\label{eq:tbqah}
\end{eqnarray} 
where $ M(\bm k) = M_z + t k^2 $ and $ \bm \sigma=(\sigma^x,\sigma^y,\sigma^z) $ are Pauli matrices. (They should not cause confusion with the notation of conductivity.) The constant $ \Delta_\text{sc} $ is the s-wave SC order parameter, $ t $ is hopping, $ A $ is spin-orbit coupling and $ M_z $ is magnetization. Depending on $ M_z, \Delta_\text{sc}$ and  $\mu $, this model has three topologically distinct phases. A TSC with single chiral Majorana edge state is realized when $\mu^2+\Delta_\text{sc}^2 > M_z^2   $. When $\mu^2+\Delta_\text{sc}^2 < M_z^2   $, it is a QAH insulator if $ M_z A <0  $, and a trivial insulator if $ M_z A >0 $ \cite{Qi2010}.

\begin{figure}
	\centering
	\includegraphics[width=0.9\linewidth]{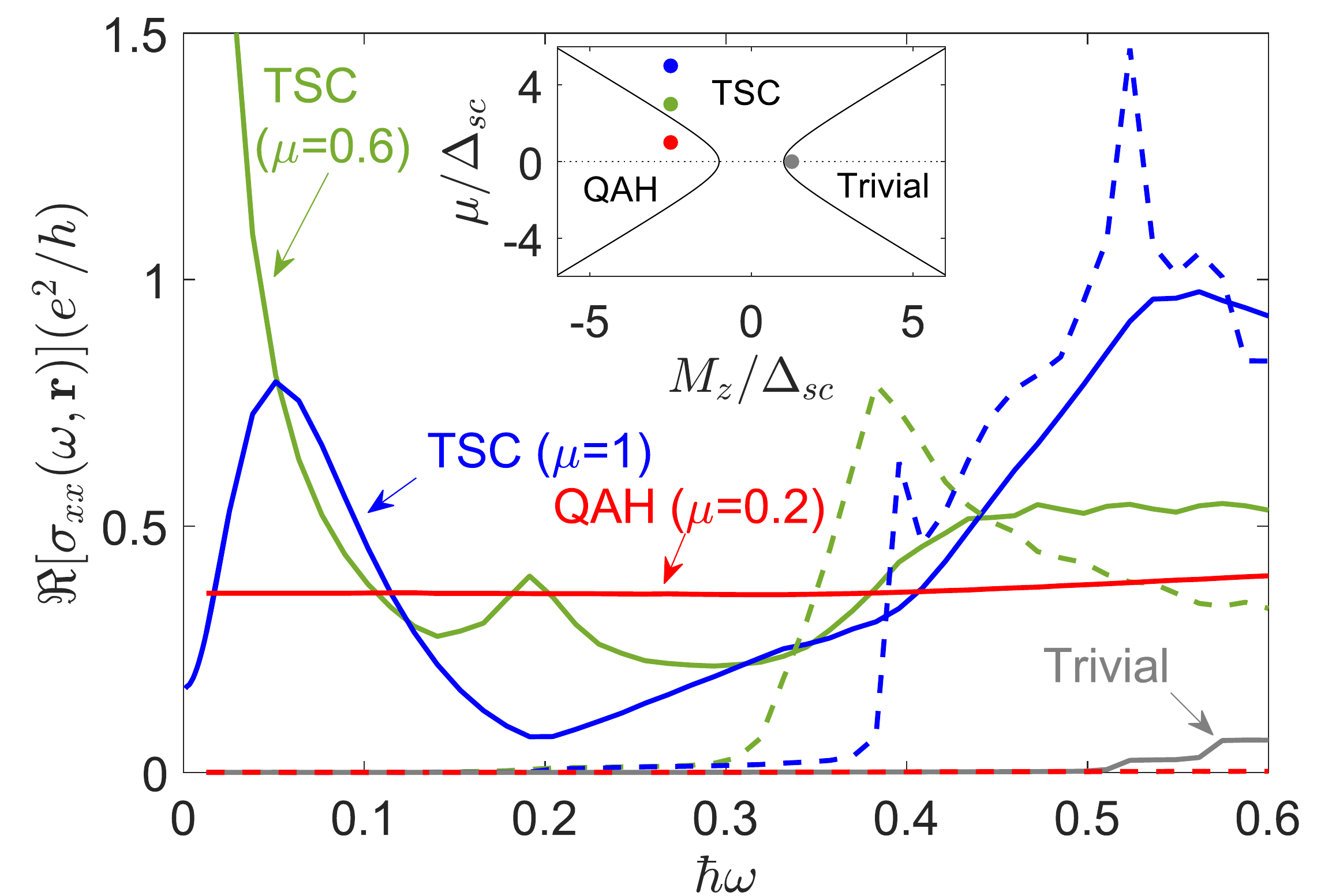}
	\caption{{The real part of the optical conductivity across the phase transition from a quantum anomalous Hall (QAH) insulator ($ \mu=0.2 $) to a topological superconductor (TSC) with single Majorana edge mode ($ \mu=0.6 $ and $ \mu=1 $). All the three cases have $ M_z=-0.5 $.  Solid curves are results on the edge and dashed ones are for the bulk.  
	The grey curve is for a trivial insulating phase with $ \mu=\Delta_\text{sc}=0 $ and $ M_z=0.25 $. Other parameters in Eq.(\ref{eq:tbqah}) are $ t=A=1 $ (regarded as the energy unit),  $ \Delta_\text{sc}=0.2  $, and $ k_B T = 10^{-3}$. 
	The dots in the inset,  with colors corresponding to the curves, shows the positions of  chosen parameters in the topological phase diagram.}}
	\label{fig:qahmus}
\end{figure}

\begin{figure}
	\centering
	\includegraphics[width=0.9\linewidth]{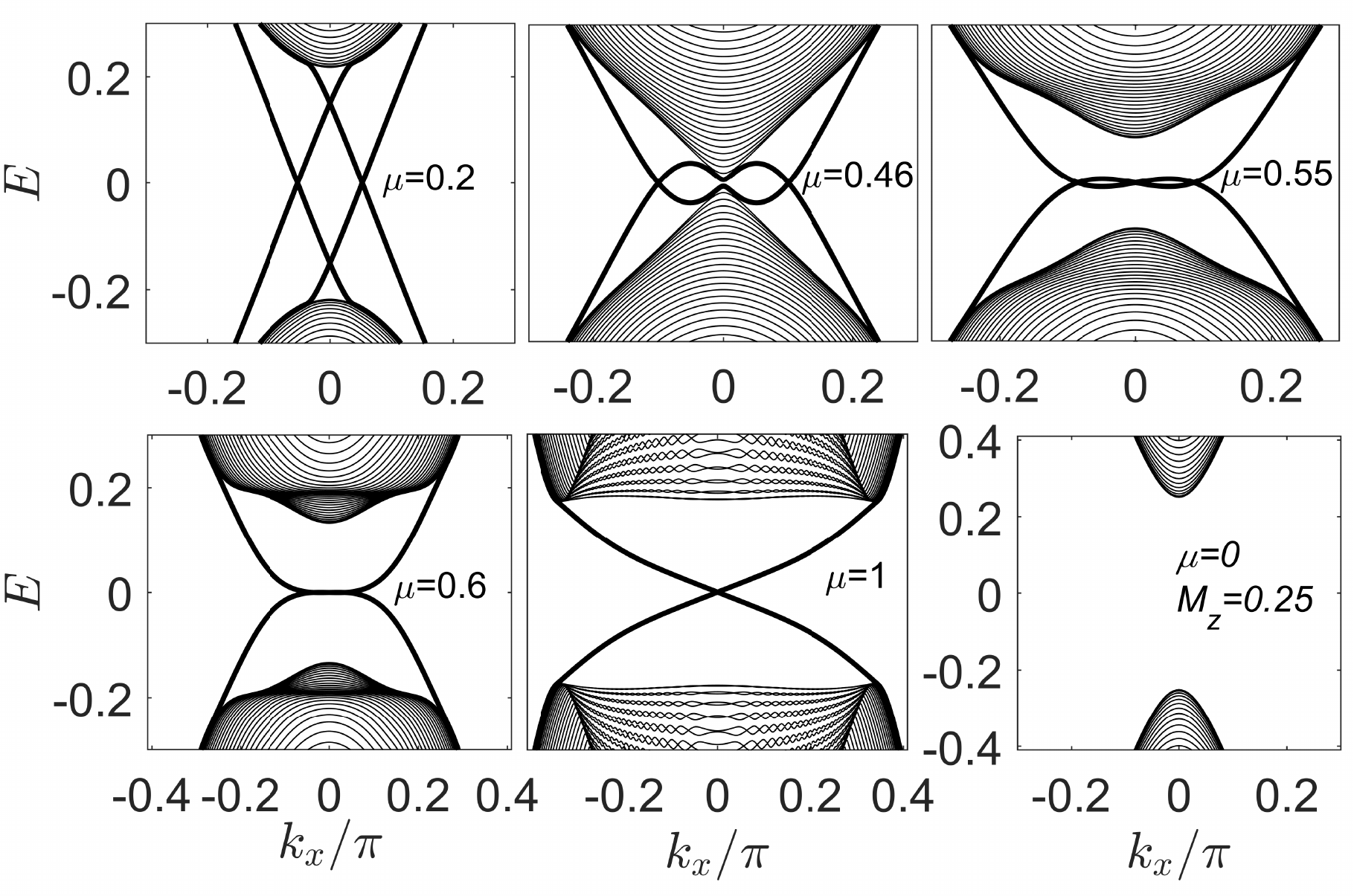}
	\caption{The evolution of band structures as the chemical potential $ \mu $  changes, obtained from Eq.(\ref{eq:tbqah}). The boundaries in the y-direction are open. Note that there are two chiral Majorana modes corresponding to the two edges. Also, the redundant degrees of freedom (the Majorana modes at $ k<0 $) are present here which are not independent. 
	}
	\label{fig:bandstructures}
\end{figure}

By varying $ \mu $ and keeping $ M_z $ and $ \Delta_\text{sc} $ unchanged, we can drive the system from a QAH phase to a TSC that has a single chiral Majorana edge mode. The local optical conductivities for three typical values of $ \mu $ is shown in FIG.  \ref{fig:qahmus}. When it is a QAH insulator ($ \mu=0.2 $), the optical conductivity on the edge is almost an $ \omega $-independent constant. 
The value is lower than $ e^2/2h $ because the spot size along the transverse direction ($ Y $) is not large and only part of the edge state is covered.  
In the TSC phase, there are two typical kinds of curves. One of them has a peak at $ \omega=0 $ while the other has a peak at finite $ \omega $. This  is due to different dispersion relations of the Majorana edge states as shown in FIG. \ref{fig:bandstructures}. When $ \mu $ is above and close to the critical value $ \mu_c=\sqrt{M_z^2-\Delta_\text{sc}^2}=0.46 $, the dispersion of the edge state is not monotonic. In fact, the spectrum crosses zero-energy three times. As $ \mu $ increases, the edge dispersion becomes monotonic after a Lifshitz transition at  $ \mu=0.6 $. Near this point the edge modes becomes very flat. The flat dispersion results in a divergent density of states at zero energy and thus a peak of $\Re[ \sigma_{xx}(\omega,\bm r)] $ appears at zero frequency. 
This peak moves to a higher frequency as $ \mu $ further increases and the dispersion becomes more and more linear (say at $ \mu=1 $). Then, it starts to look similar to the results of previous models. 
The results for a trivial insulating phase ($ M_z=0.25, \Delta_\text{sc}=\mu=0 $) are also shown in FIG. \ref{fig:qahmus} for comparison. The optical conductivity vanishes when the frequency is lower than the insulating gap (around $ 0.5 $). Above that, it becomes non-zero. The smallness of $ \Re[\sigma] $ is due to the small density of states (because of numerical finite-size effect) near the gap, as seen in  FIG. \ref{fig:bandstructures}. 

\paragraph {Conclusion and discussion --- }

We have shown that chiral Majorana edge states in TSCs can be detected by measuring the local optical conductivity. Compared to normal edge states, the signals of Majorana fermions is comparable or even stronger, and it shows 
qualitatively distinct features such as the frequency and temperature dependencies , i.e. $ \Re[\sigma_{xx}]_{T=0} \sim \omega^2 $ and $ \Re[\sigma_{xx}]_{\omega=0} \sim T^2 $, for small $ \omega $ and $ T $. \cite{supp} 
Also, near the  topological phase transition from a QAH insulator to a TSC, the Majorana fermions have rather flat energy dispersion and the low-frequency optical response becomes gigantic.

Only the real part of the optical conductivity $ \Re[\sigma(\omega)] $ is discussed here. The imaginary part $ \Im[\sigma(\omega)] $ may also be measured with optical microscopy methods. However, in superconductors,  a purely imaginary diamagnetic term, $ i \frac{n_se^2}{m\omega} $, always appears. 
In some circumstances it may be used to distinguish $ p $-wave superconductors from conventional ones \cite{Asano,Bakurskiy}. But in our case, it surges up at the low-frequency limit and thus not really informative about the Majorana edge states.

One way of realizing the chiral Majorana modes described by our theoretical models is to use the surface states of topological insulators such as  Bi$ _2 $Se$ _3 $ \cite{supp}. In this case, the chemical potential $ \mu $ should be inside the surface magnetization gap ($ \Delta_m \sim 50meV $ \cite{Tokura2019,Ko2020} ). Assuming $ \mu=50meV $, the SC gap $ \Delta_{sc}=0.1meV$ and the detection spot size $ d=5nm $ \cite{Lee_SciAdv2020}, we estimate that the optical conductivity has a maximum value of $ \Re[\sigma_{xx}(\omega_{0})] \approx 10e^2/h $ at the peak position $ \hbar \omega_{0} \approx  0.003 meV  $, or $ \omega_{0} \approx 4.5 $ GHz.

\begin{acknowledgments}
	
	We thank Yoshinori Tokura, Zhi-Xun Shen and Tian Liang for helpful discussions. N.N. was supported by Ministry of Education, Culture, Sports, Science, and Technology Nos. JP24224009 and JP26103006, JSPS KAKENHI Grant numbers 18H03676 and 26103006, and Core Research for Evolutionary Science and Technology (CREST) No. JPMJCR16F1 and No. JPMJCR1874, Japan. Y.T. was supported by Grant-in-Aid for Scientific Research on Innovative Areas, Topological Material Science (Grants No. JP15H05851, No. JP15H05853, and	No. JP15K21717) and Grant-in-Aid for Scientific Research B (Grant No. JP18H01176) from the Ministry of Education, Culture, Sports, Science, and Technology, Japan (MEXT). Y.T. was also supported by Grant-in-Aid for Scientific Research A (KAKENHI Grant No. JP20H00131) and the JSPS Core-to-Core program Oxide Superspin International Network. J.J.H. was supported by RIKEN Incentive Research Projects.
	
\end{acknowledgments}

\bibliographystyle{apsrev4-1}

\begin{thebibliography}{99}	
\bibitem{Kitaev2001} A. Y. Kitaev, Physics-Uspekhi { 44}, 131 (2001).
\bibitem{Wilczek} F. Wilczek, Nat. Phys. {  5}, 614 (2009).
\bibitem{Sato} M. Sato and Y. Ando, Rep. Prog. Phys. 80 076501 (2017).
\bibitem{Tanaka_JSPS} Y. Tanaka, M. Sato and N. Nagaosa, J. Phys. Soc. Jpn. 81, 011013 (2012).

\bibitem{Read} N. Read and D. Green, Phys. Rev. B 61, 10267 (2000).

\bibitem{Fu2008} L. Fu and C. L. Kane, Phys. Rev. Lett. {  100}, 096407 (2008).
	
	
\bibitem{Qi2010} X. L. Qi, T. L. Hughes and S. C. Zhang, Phys. Rev. B {  82}, 184516 (2010).

\bibitem{Chung2011} S. B. Chung, X. L. Qi, J. Maciejko and S. C. Zhang, Phys. Rev. B {  83}, 100512(R) (2011).
		
\bibitem{JJH_CP2019} J. J. He, T.  Liang, Y. Tanaka and N. Nagaosa, Commun. Phys. 2, 149 (2019). 
	
\bibitem{Vic2009} K. T. Law, P. A. Lee and T. K. Ng, Phys. Rev. Lett. {  103}, 237001 (2009).

\bibitem{Lunhui} L. -H. Hu, C. Li, D. -H. Xu, Y. Zhou and F. -C.  Zhang, Phys. Rev. B {  94}, 224501 (2016).


\bibitem{Machida} T. Machida et al. Nature Materials  18, 811 (2019).

\bibitem{Yazdani2014} S. Nadj-Perge, I. K. Drozdov, J. Li, H. Chen, S. Jeon, J. Seo, A. H. MacDonald, B. A. Bernevig, and A. Yazdani, Science {  346}, 602 (2014).
\bibitem{JiaPRL} H. -H. Sun, et al., Phys. Rev. Lett. {  116}, 257003 (2016).
	
	\bibitem{Chiu} C. K. Chiu et al., Sci. Adv. 6: eaay0443 (2020).

\bibitem{Kouwenhoven2018} H. Zhang et al., Nature   556, 74 (2018).

\bibitem{JWang2015} J. Wang, Q. Zhou, B. Lian and S. C. Zhang, Phys. Rev. B {  92}, 064520 (2015).
	
	
\bibitem{Qinglin2017} Q. L. He et al., Science {  357}, 294 (2017).
	
	
	

	\bibitem{Wen2018} W. Ji and X. -G. Wen, Phys. Rev. Lett. {  120}, 107002 (2018).
	
	\bibitem{Sau2018} Y. Huang, F. Setiawan and J. D. Sau, Phys. Rev. B {  97}, 100501(R) (2018).
	
	
	\bibitem{ZWang} Z. Wang et al. Science 367, 104-108 (2020).  
	
	\bibitem{Lai} K. Lai, W. Kundhikanjana, M. A. Kelly, Z.-X. Shen, J. Shabani, and M. Shayegan,
 Phys. Rev. Lett. 107, 176809 (2011).
	 
	
	\bibitem{Shen2} M. Allen, Y.-T. Cui, E. Y. Ma, M. Mogi, M. Kawamura, I. C. Fulga, D. Goldhaber-Gordon, Y. Tokura, and Z.-X. Shen,
	Proc. Natl. Acad. Sci. 116, 14511-14515 (2019).
	
	
	
	\bibitem{NN1995} N. Nagaosa and M. Kohmoto
	Phys. Rev. Lett. 75, 4294 (1995).
	
	\bibitem{Fukui} T. Fukui, K. Shiozaki, T. Fujiwara and S. Fujimoto, J. Phys. Soc. Jpn. 81, 114602 (2012);
	
	\bibitem{supp} See Supplemental Material at 
	http://link.aps.org/supplemental/10.1103/PhysRevLett.
	126.237002 
	for details
	of derivation, additional information on the temperature
	effects, and discussion about the results on surfaces of
	topological insulators. 
	
	\bibitem{Furusaki} A. Furusaki, M. Matsumoto and M. Sigrist, Phys. Rev. B 64, 054514 (2001).
	
	\bibitem{Lee_SciAdv2020} K. Lee, M. I. B. Utama, S. Kahn, A. Samudrala, N. Leconte, B. Yang, S. Wang, K. Watanabe, T. Taniguchi, M. V. P. Altoé, G. Zhang, A. Weber-Bargioni, M. Crommie, P. D. Ashby, J. Jung, F. Wang, and A. Zettl, Science Advances 6, eabd1919 (2020).
	
	\bibitem{Asano} Y. Asano, A. A. Golubov, Y. V. Fominov  and Y. Tanaka, 
	Phys. Rev. Lett. 107, 087001 (2011).
	
	\bibitem{Bakurskiy} S. V. Bakurskiy, Ya. V. Fominov, A. F. Shevchun, Y. Asano, Y. Tanaka, M. Yu. Kupriyanov, A. A. Golubov, M. R. Trunin, H. Kashiwaya, S. Kashiwaya  and Y. Maeno
	Phys. Rev. B 98, 134508 (2018).
		
    \bibitem{Tokura2019}  Y. Tokura, K. Yasuda, and A. Tsukazaki, Nat. Rev. Phys. 1, 126 (2019).
	
	\bibitem{Ko2020} W. Ko, M. Kolmer, J. Yan, A. D. Pham, M. Fu, F. Lüpke, S. Okamoto, Z. Gai, P. Ganesh, and A.-P. Li, Phys. Rev. B 102, 115402 (2020).
	
	
	
\end{thebibliography}

\end{document}